# Axis2UNO: Web Services Enabled Openoffice.org

B.A.N.M. Bambarasinghe, H.M.S. Huruggamuwa, R.G. Ragel and S. Radhakrishnan
Department of Computer Engineering, Faculty of Engineering,
University of Peradeniya CP 20400 Sri Lanka.
madhurajith@gmail.com, shivanthah@gmail.com, roshanr@pdn.ac.lk, swarnar@pdn.ac.lk

*Abstract—* **Openoffice.org is a popular, free and open source office product. This product is used by millions of people and developed, maintained and extended by thousands of developers worldwide. Playing a dominant role in the web, web services technology is serving millions of people every day. Axis2 is one of the most popular, free and open source web service engines. The framework presented in this paper, Axis2UNO, a combination of such two technologies is capable of making a new era in office environment. Two other attempts to enhance web services functionality in office products are** *Excel Web Services* **and** *UNO Web Service Proxy.* *Excel Web Services* **is combined with** *Microsoft SharePoint* **technology and exposes information sharing in a different perspective within the proprietary** *Microsoft* **office products.** *UNO Web Service Proxy* **is implemented with** *Java Web Services Developer Pack* **and enables basic web services related functionality in Openoffice.org. However, the work presented here is the first one to combine** *Openoffice.org* **and** *Axis2* **and we expect it to outperform the other efforts with the community involvement and feature richness in those products.**

## I. INTRODUCTION

ONE of the current trends in information technology is decentralized computing - the opposite of centralized computing which was prevalent during the early days of computers. Decentralized computing is the allocation of both hardware and software resources to each individual computing system [1]. Web services play a vital role in this situation and allow user environment to independently access to a whole set of vital and interesting services all around the world [3]. Typically, a web service is just an API (Application Programming Interface) that can be accessed over a network, such as the *Internet*, and can be executed on a remote system where the requested service is hosted [6].

To access web services usually customized software such as software from various service providers to their clients to access some providers' data or pre programmed web sites such as which have concerns about the security, are used. However for the general office use, data from a remote web service is still little cumbersome because office software packages do not provide sufficient access to web services up to now.

So opening of doors and windows for web services in *Openoffice.org* is going to expose a completely new world of information easily available and accessible in the office environment. However implementing a whole set of functionalities needed in accessing web services is really a waste as there are already implemented open source web service engines that could be utilized on this endeavour. The role of a web service engine is to accept requests and related parameters to web services, executing the web service and finally forwarding the outcome to the requesting program. In addition to this, web service engines will also provide some services on security, data formatting, etc. The web service engine we chose to use in this project, *Axis* [4], is one of the popular web service engines. Other web service engines are implemented for *Java* and *.NET* development by corresponding vendors.

*Openoffice.org* is a free and open source [7] office package and there are thousands of developers extending the functionality of this package. If one want to incorporate the web services functionality in her component she has to work all from the scratch as there is no integrated web service engine available in *Openoffice.org*. It is truly tedious and repetitive if it is needed to implement the web service functionality, all the time when you develop an office component. Therefore, the goal of this project is to *broaden the extensibility of Openoffice.org by integrating Axis2 web services engine into Openoffice.org 2.0 as a component that can be accessed by any developer.* By this integration, we will be able to develop an *Openoffice.org* specific and more object oriented web service interface so that the development is more intuitive. In addition, all the advantages such as speed, security and flexibility when using *Axis2* for web service access is available right inside *Openoffice.org* when this developed component is in the developers' toolbox.

The remainder of this paper is organized as follows. We present a summary of relevant past work in Section II. In Section III we describe the background of this project - *Openoffice.org* and web services with *Axis2*. Section IV of this paper outlines the software design aspects and Section V details software implementation. We present the evaluation method in Section VI and conclude in Section VII.

## II. RELATED WORK

Web services have been identified as one of the valuable components in this era of information technology. Therefore, they are deployed in almost everywhere with the basic usage of exploiting it to accomplish a certain task in an application. The application that uses web services could be either a web application or a desktop application.

There are a number of web applications, which make use of web services. One of the major reasons that web applications use web services is that they can encapsulate important data behind a web service and expose only display logic in the web application. This method is far secured than directly accessing databases from the web application. Some of the well-known web service providers are *Google* and *Amazon* where they provide web service interface for their traditional services like search the Web or querying the online bookstore.

Desktop applications have enough reasons to use web services to accomplish their task. For example, a desktop application used for stock broking needs to check out the stock market data. This task can be accomplished in that particular application by invoking a web service provided by the stock market. Then the application does not need to access the stock market database directly and since the web services use the http (*hypertext transfer protocol*), the information can be transferred even through firewalls.

With this much of vast advantages Microsoft Office 2007 package has taken some step to accommodate web services functionality into its *Excel* product. This setup is called *Excel Web Services* [10]. *Excel Web Services* can consume a web service hosted in a Microsoft SharePoint server. In addition to hosting web services the SharePoint provides other services such as portals, collaboration, etc. Since *Excel* has been integrated with SharePoint technology, it is composed with a number of valuable features regarding remote services and data.

A RESTful type web service can be invoked inside *Openoffice.org* using *Openoffice.org* macros [11]. Macros can make use of existing file access and XML parser UNO services to invoke such a RESTful web service. Macros usage is fairly limited as those cannot be used to invoked SOAP/WSDL (Web Services Description Language) type web services and when more functionality is needed some other method should be exercised.

*Java Web Services Developer Pack* (JWSDP) and *Openoffice.org* combination has gone some distance in terms of research and development of allowing web services in *Openoffice.org* [9]. This allows dynamic binding of web services using WSDL descriptions. Even though this allows a number of functionalities including complex data types handling, it still lags behind against the functionalities of *Axis2* because of *Axis2's* unique features, such as support for different message formats, fast and pluggable deployers, axiom object model, etc. [4]

As described in this paper, enabling web services using *Axis2* is a more intuitive approach for accessing web services. Although each web service platform contains all the basic tools for consuming majority of available services, each has its own unique features. Developers who expertise with such platforms master these special features to accomplish some specific tasks. In this background, embedding *Axis2* inside *Openoffice.org* will make an easy entry for thousands of *Axis2* mastered developers into *Openoffice.org* development, showing the power of two mostly used technologies. *Axis2* and related products are totally focused on the web services technology. Therefore, there are a number of useful functionalities that can be exposed including some properties of security and scalability.

The project drives web services developers of office products towards an object-oriented approach. Finally, the *Axis2UNO* presented in this paper is targeted to become the best choice for web services in the *Openoffice.org* UNO environment. This target is motivated by the large number of developers for *Axis2* and its popularity.

## III. OPENOFFICE.ORG AND WEB SERVICES

*Openoffice.org* was established when *Sun Microsystems Inc.* released the source code of their popular office package Staroffice to the open source world in year 2000 [2]. Then in spring 2002 Staroffice 6.0 and *Openoffice.org* 1.0 both were released sharing the same code base.

*Openoffice.org* SDK is the software development kit produced by the community to enhance the functionality of the *Openoffice.org* software package. It contains header files, necessary tools and documentation. *Openoffice.org* architecture makes the development and extension of any functionality, language independent by introducing open office specific interface definition language called Universal Network Object Interface Definition Language (UNOIDL).

UNOIDL is used to define interfaces and it is used by the *Openoffice.org* to identify services internally. Interfaces declared using UNOIDL are needed to be converted in to a high level programming language so that the functionality can be implemented. Tools needed for converting UNOIDL into high level languages are provided in the *Openoffice.org* SDK (Software Development Kit).

*Openoffice.org* package is a collection of individual programs executing together to provide necessary functionality as an office software package. This environment is called the UNO (Universal Network Objects) environment. These individual programs are called Services in *Openoffice.org* terms. A service may contain *Properties*, *Functions*, *Enumerations*, etc. These properties, functions, and enumerations are coinciding with high level programming language equivalents. The tools provided within the SDK perform a *one to one* mapping when converting UNOIDL into other high level programming languages.

A service created using the *Openoffice.org* SDK can be installed in the UNO environment so that this service is accessible later by any other services or add-ons (user interface components that are used to provide some service to the end user) by querying into the *Service Manager*. The service manager is the program (UNO service) inside the *Openoffice.org* package, which takes care of instantiating services on behalf of other services. For this purpose it maintains a registry of all installed services and when a request is received, it looks for the actual service and makes an instance of that service and passes its reference to the requesting service or add-on. Figure 1 shows the UNO environment where the service manager is a service, which does the instantiation of other UNO services.

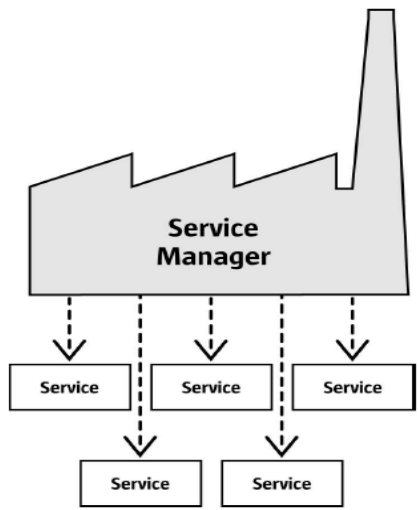

Figure 1. Openoffice.org UNO environment (taken from [2])

The definition for *web services* given by *World Wide Web Consortium* [5] which establishes the standards for web services is,

"*A web service is a software system identified by a URI whose public interfaces and bindings are defined and described using XML. Its definition can be discovered by other software systems. These systems may then interact with the Web service in a manner prescribed by its definition, using XML-based messages conveyed by Internet protocols.*"

Web services operate on the Service Oriented Architecture (SOA). An SOA is an architectural style used by computer systems for creating and using business processes, packaged as services, throughout their lifecycle [8]. To achieve much of interoperability, web services are standardized to provide:
- a standard mark-up language for communication,
- a common message format for exchanging information,
- a common service specification format, and
- a common means for service lookup.

Apache *Axis2* [4] is a third generation web services engine built on top of a highly modular architecture. Design of *Axis2* has taken few facts into consideration and most of these aims high scalability and improved performance of the system. It is used as the middleware of both client side and server side in many scenarios. This *Axis2* middleware takes care of web services related work such as SOAP protocol handling, data binding, etc. Figure 2 depicts the behaviour of the *Axis2* web services engine.

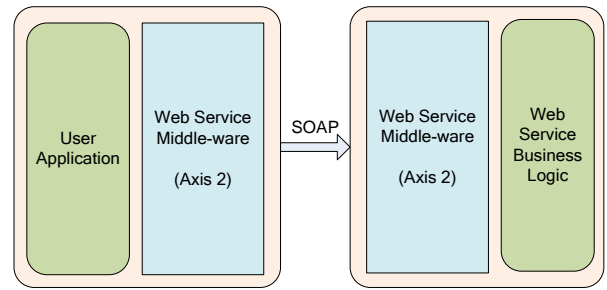

Figure 2. Apache *Axis2* Architecture

Some of the key concepts include
- Support for both synchronous and asynchronous interactions
- Support for multiple transport protocols such as TCP, HTTP, SMTP, JMS
- Embeddable in a standalone application or can be packaged in a *web archive* file so that giving the end user the flexibility in deciding the deployment and packaging options.
- Hot deployment facility with a new archiving format
- Axiom object model to improve performance
- Concept of Handlers and Phases to improve scalability. *Handlers* can be thought of as a pipeline that passes the message context through. A *phase* is a logical grouping of handlers and contains rules such as "*Phase First*",

"*Phase Last*", "*Before*", and "*After*", which defines a very flexible execution framework. Each handler is responsible for processing a certain aspect of the SOAP message. For example, the Addressing handlers deal with the WS-Addressing headers in the SOAP message
- Allowing user to support message exchange pattern (MEP) and avoiding *Axis2* assuming the MEP. Currently supported MEPs are IN_ONLY and IN_OUT.

In the open source world, *Axis2* is the choice for web services with its scalability and performance through the key concepts described above. *Axis2* is available in Java and C implementations and we chose to use the Java implementation. The following section will show how the system is designed and will highlight the software architecture of the system.

## IV. SOFTWARE DESIGN

There were two major requirements identified under the software design of this project and they are: (1) exposing the web services functionality inside *Openoffice.org* UNO environment, and (2) encapsulating web services related contents (such as WSDL, data type conversion) as much as possible inside a UNO service so that the other UNO services need not consider web services specific content.

The service or the add-in who is going to use web services should have a mechanism to pass data needed by the remote web service and it should receive the response from the remote web service. For this purpose, the *Axis2WebServices* sub system is used as a middleware.

After adding web services into the UNO environment, existing services, web services designed by us and the service manager are arranged in the way as depicted in Figure 3. As depicted, UNO services or add-ons which need the web services instantiate them by using the service manager.

In reality a web service is invoked using a web service request. A web service request consists of several input/output parameters, name of the web service, security parameters, etc and exists in the form of a XML document. In the UNO environment *Axis2RequestUNO, Axis2ParameterUNO and Axis2UNO* represent corresponding components and other services outside the *Axis2WebServicesUNO* sub system use instances of *Axis2RequestUNO* and *Axis2ParameterUNO* to communicate with the *Axis2WebServiceUNO*.

*Axis2ParameterUNO* represents an input or output parameter passed to or returned by the web service in the UNO environment. These parameters are used by the *Axis2WebServices* to build the *SOAP* request and as the means of forwarding response of remote web services to the requesting UNO services. *Axis2ParameterUNO* consists of name, value, namespace, etc.

The UNO service named *"Axis2UNO"* wraps the basic functionality of the *Axis2* web services engine. It parses the gap between two technologies by allowing the *Axis2UNO* to be initiated using the service manager and allowing other UNO services to communicate with it. When a request is received by the *Axis2WebServices,* it binds a remote web service internally using the data available in the request. On a request to execute the remote web service the *Axis2UNO* service generate the SOAP request, execute the remote web service and acquire the response. Then with this response, return parameters are initialized and forwarded to the requested UNO service.

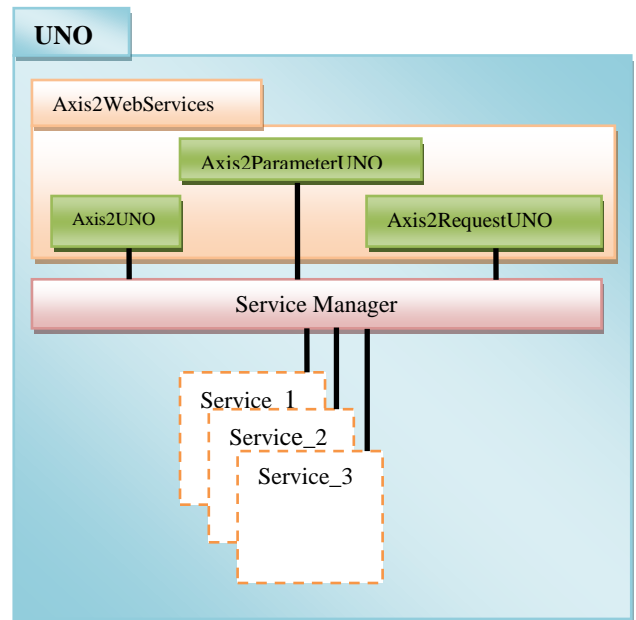

Figure 3. UNO Web Services Architecture

## V. IMPLEMENTATION

In this section, we explain how the software design proposed in the previous section is implemented in our project. The development of the UNO services was performed using the tools available in *Openoffice.org* SDK. The *Axis2WebServices* sub system was implemented using Java. Netbeans IDE was used as the Java development environment as it provides a plug-in that enables invoking *Openoffice.org* SDK tools in the IDE itself.

First, the interfaces are defined in UNOIDL as the UNO environment will only understand UNOIDL definitions. Those definitions are installed with the concrete implementation later. The *Service Manager* uses these interface definitions to choose the services to instantiate on a request. These definitions should be converted to a high level programming language so that the functionality can be implemented. One UNOIDL definition can have many concrete implementations and the service manager decides which implementation to instantiate. There the service manager behaves as a factory in software engineering terms. In order to convert UNOIDL definitions to high level languages, *Openoffice.org* SDK provides several compilers for each language. Basically compilers are available for C and Java. We invoked *UNOIDL Java compiler* through Netbeans.

The *Axis2UNO* interface, that allows instantiating an Axis2UNO using a given *Axis2RequestUNO*, is depicted in Figure 4. The remote web service can be executed depending on the expected communication pattern with the remote web service. If it is intended to receive a value from the web service it should be executed using *OutInExecute()* or else it should be executed using *outExecute()*. In a *outInExecute()* return values are received by the requesting UNO service as a array of *Axis2ParametersUNO*. There the requesting web service is free of XML processing as it is done inside the *Axis2WebService* UNO service.

```
Interface Axis2UNO {
    public Axis2WebService (Axis2RequestUNO);
    public List< Axis2ParameterUNO> outInExecute();
    public boolean outExecute();
}
```
Figure 4. Axis2UNO Interface

Figure 5 depicts the interface of Axis2RequestUNO class. A web service request should have input and output parameters. Accessor functions (Get/Set) are provided for those parameters. The other methods define the web service related activities.

```
Interface Axis2RequestUNO {
    public addParameter(Axis2ParameterUNO);
    public removeParameter(String name);
    public getParameter(String name);

    public addReturnParameter(Axis2ParameterUNO);
    public removeReturnParameter(String name);
    public getReturnParameter(String name);

    public void setOperation(String);
    public void setNamespace(String);
    public void setPrefix(String);
    public void setAction(String);
    public void setExceptionOnSOAPFault();
}
```
Figure 5. Axis2RequestUNO Interface

The Axis2ParameterUNO is just a container with the following data:

- String *name*
    Name of the parameter
- Void *value*
    Value of the parameter
- String *namespace*
    Each input or output variable of a web service has a namespace this variable defines that namespace.
- Boolean *nullable*
    A flag to represent whether it can be null or not
- Integer *maxOccurs*
    Maximum occurance of the parameter
- Integer minOccurs
    Minimum occurance of the parameter

All that data can be publically accessible with their variable names.

VI. EVALUATION

The newly implemented functionality is accessed by a user interface extension add-in and data extracted from the web service is directly put on an office document such as a spreadsheet. An example add-in is the stock market data extractor. It asks the user for interested instruments and fills a spreadsheet with extracted data. A sample is depicted in Figure 6, and the add-in that performs the work in Figure 7.

Figure 6. Spreadsheet with Market Data

Typically, this kind of a task is accomplished by the following procedure:

1. Create a *Axis2RequestUNO* service using *Openoffice.org* service manager
2. Set namespace of the web service
3. Set action of the web service
4. Set input and output parameters for the web service creating new Axis2ParameterUNO using *Openoffice.org* service manager.
5. Create *Axis2UNO* service using *Openoffice.org* service manager
6. Execute the web service
7. Process output of the web service and put data in the office document

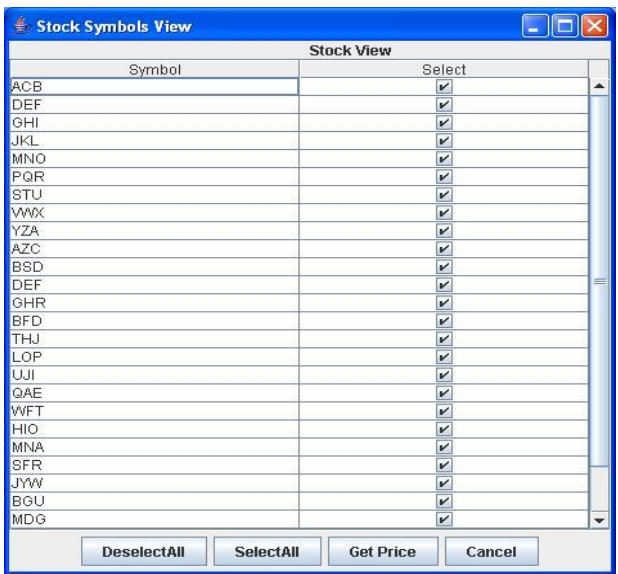

Figure 7. Market Data Extractor Add-in User Interface

Accessing an object from the service manager is the typical procedure. With this approach any future developments do not need to struggle with the intricacies of *Axis2* API. Web services are accessed with the routine *Openoffice.org* development procedure.

## VII. CONCLUSION

In this paper, we presented *Axis2UNO*, the first project of this kind that enables web services in the infamous open-source office product *Openoffice.org*. *Axis2UNO* combines *Axis2,* a popular web services engine with *Openoffice.org*. This combination exposed an object oriented API to the developers by using the Java implementation of *Axis2*.

Even with the minimal web services functionality exposed through the model, the outcome is fascinating. However, a number of future works that could reveal the full power of web services in *Openoffice.org* are identified. The identified future works are:

- The web service UNO component can be expanded such that the vast functionality of Apache *Axis2* is available in *Openoffice.org*.
- With the proposed evaluation model, add-in components for only specified web services can be accessed. But it is too limited. There should be a robust mechanism to access any web service and put the response in an office document. For that an implementation on top of a good policy is required.
- The most vital and essential requirement is to have the ability of saving web service information in office documents and accessing those web services automatically when required. This can be accomplished by creating new document types and loaders in *Openoffice.org*.